\documentclass[preprint,aps,prb]{revtex4}

\newcommand{\Ref}[1]{Ref.~\onlinecite{#1}}

\def\eb{\begin{equation}}   
\def\ee{\end{equation}}     

\def\s{\sin\theta}

\def\eq#1{Eq.~(\ref{#1})}
\def\eqs#1#2{Eqs.~(\ref{#1}) and (\ref{#2})}

\def\xv0{{\vec{x}}^{\,0}}
\def\xdot{\dot{x}}
\def\xddot{\ddot{x}}

\def\s0{\sigma_{0}}

\begin{document}

\author{Jeremy Schiff}
\email{schiff@math.biu.ac.il}
\affiliation{Department of Mathematics, Bar-Ilan University,\\
              Ramat Gan 52900, Israel}
\author{Bill Poirier}
\email{Bill.Poirier@ttu.edu}
\affiliation{Department of Chemistry and Biochemistry, and
         Department of Physics, \\
          Texas Tech University, Box 41061,
         Lubbock, Texas 79409-1061}

\title{Quantum Mechanics Without Wavefunctions}

\begin{abstract}

We present a self-contained formulation of spin-free nonrelativistic quantum 
mechanics that {\em makes no use of wavefunctions or complex amplitudes of 
any kind}.  Quantum states are represented as  ensembles of real-valued quantum 
trajectories, obtained by extremizing an action and satisfying energy conservation.
The theory applies for arbitrary configuration spaces and system 
dimensionalities.   Various beneficial ramifications---theoretical, computational, and
interpretational---are discussed.

\end{abstract}


\maketitle


\section{INTRODUCTION}
\label{intro}

For nearly a century, quantum mechanics has presented philosophical 
and interpretational conundrums that remain as controversial as  ever.  
Far from disappearing into the realm of esoteric academic debate, recent 
experimental  advances, e.g. in entanglement, decoherence, and quantum 
computing, have brought such questions to the forefront of topical interest. 
The various competing viewpoints---Copenhagen,\cite{vonneumann}
Bohmian,\cite{bohm52a,holland,wyatt} Many Worlds,\cite{everett57} etc.---differ 
substantially in terms of their ontological interpretation of  $\Psi$ and its collapse,
yet are all alike in their reliance upon a complex-valued wavefunction, and its 
propagation via the time-independent/dependent Schr\"odinger equation (TI/DSE).
Evidently, a completely self-contained, classical-like, and real-valued formulation 
of quantum theory {\em without} wavefunctions---if such a thing were possible---would 
necessarily present novel and potentially very important interpretational and 
computational ramifications.  

The aim of this paper is to show that for non-relativistic spin-free quantum mechanics, 
the above {\em can} be realized (A different approach is described in \Ref{gonzalez08}.) 
In particular, a quantum state can be represented
as an ensemble of real-valued quantum trajectories, satisfying a self-contained partial 
differential equation (PDE). This was shown in a recent paper\cite{poirier10nowave} 
by one of the authors (Poirier) for both the one-dimensional (1D) TISE and TDSE.  
It transpired that similar work had already been done for the TISE in 1D,\cite{bouda03}
the TDSE in 3D,\cite{holland05} and in greater generality.\cite{holland05b} 
In the current paper, we simplify, unify, and generalize these previous constructions, 
presenting quantum trajectory PDEs for arbitrary configuration spaces and system 
dimensionalities. The goal here is 
to present an alternative, standalone {\em reformulation} of quantum mechanics, 
that neither relies on the TDSE nor makes any mention of any external constructs 
such as $\Psi$, and which, in addition, is likely to provide  
 far-reaching benefits for numerical calculations, e.g. of accurate
quantum scattering dynamics for chemically reactive molecular systems. 

The relevant PDEs can be derived from an action principle. We discuss 
symmetries of the action principle, associated conserved quantities, and other 
properties, such as a heretofore unexpected Hamiltonian structure in the case of 
1D time-independent quantum mechanics (TIQM), which also serves as the basis 
of an accurate many-D numerical scheme. We single out a specific choice of 
Lagrangian or gauge, and show why this choice may be regarded as physically 
preferred.  Likewise, we single out a specific choice of trajectory-labeling 
coordinate(s), in terms of which the resultant PDE exhibits no explicit coordinate 
dependences.
 We present simple analytical solutions, and discuss some 
initial numerical results that appear very promising for molecular and chemical physics 
applications. The reformulation of quantum mechanics in terms of trajectory ensembles,
in addition to shedding light on complex theoretical issues, evidently also provides 
important practical benefits.


\section{The 1D Time-independent Case}
\label{1Dtise}

As shown in \Ref{poirier10nowave}, 1D TIQM states can be 
represented uniquely with a single trajectory, $x(t)$. The 
theory is one of a broad class of dynamical laws, for which 
the Lagrangian and energy are of the form
\begin{eqnarray}
    L(x,\xdot,\xddot,\ldots) &  =  & {1\over 2} m \xdot^2 - V(x) - Q(\xdot,\xddot,\ldots)\ ,\label{1DL}\\
    E(x,\xdot,\xddot,\ldots) & =  & {1\over 2} m \xdot^2  + V(x) + Q(\xdot,\xddot,\ldots)\ .\label{1DE}
\end{eqnarray}
Equations~(\ref{1DL}) and (\ref{1DE}) are  natural generalizations of the 
well-known classical forms. The quantum correction, $Q$, is similar to the 
potential, in that it appears with opposite signs in $L$ and $E$, but is 
actually related to the TISE kinetic energy operator.\cite{wyatt}
It has a universal ``kinematic'' form, i.e. no explicit $x$ dependence.

As in classical theory, the quantum trajectories, $x(t)$, are obtained via 
extremization of the action, $S = \int L \,dt$.  Since $L$ is autonomous, the
resultant $x(t)$ solutions exhibit time-translation invariance and energy 
conservation for any choice of $Q$. However, for general $Q$, the conserved 
energy for Eq.~(\ref{1DL}) obtained via Noether's theorem does 
{\em not} take the form of Eq.~(\ref{1DE}). Requiring 
this equivalence imposes rather special conditions on $Q$, e.g., that $Q$ must be 
invariant under time rescaling. 
The simplest, well-behaved, nontrivial $Q$ giving the form of Eq.~(\ref{1DE}) is 
\eb
    Q(\xdot,\xddot,\stackrel{...}{x}) = \frac{\hbar^2}{4m}
 \left( \frac{\stackrel{...}{x}}{\dot{x}^3}  -\frac52 \frac{\ddot{x}^2}{\dot{x}^4}  \right).
 \label{Q1D} 
\ee
Here $\hbar$ is, in principle, an arbitrary positive constant.  
However, with the usual identification of $\hbar$ as Planck's constant, 
Eq.~(\ref{Q1D}) is equivalent to the 
quantum potential of Bohmian mechanics\cite{bohm52a,holland,wyatt}
for the 1D Cartesian TISE.\cite{poirier10nowave}
This is surprising, given that the expression Eq.~(\ref{Q1D}), being 
universal and kinematic, is determined entirely by the trajectory, $x(t)$.
In particular, no reference to  the wavefunction, $\Psi$, or the TISE itself,
is used in this derivation.  Our quantum trajectories are nevertheless Bohmian 
trajectories, although our formulation and interpretation are not 
at all that of Bohmian mechanics, because no $\Psi$ is involved.

Action extremization applied to Eqs.~(\ref{1DL}) and (\ref{Q1D}) yields the 
following fourth-order autonomous ODE, describing 1D TIQM quantum 
trajectories: 
\begin{equation}
 m\ddot{x}  + {\partial V(x)\over\partial x} + \frac{\hbar^2}{4m} 
   \left(  
    \frac{\stackrel{....}{x}}{\dot{x}^4} 
   - 8  \frac{\stackrel{...}{x}\ddot{x}}{\dot{x}^5} 
   + 10  \frac{\ddot{x}^3}{\dot{x}^6} 
   \right)
    = 0 \ . 
\label{modnewt1} 
\end{equation}
Two of the four constants of integration 
correspond to time-translation invariance and 
energy, as in the classical case. The other two constants 
determine which particular TIQM state the trajectory is associated with. 
A one-to-one correspondence thus exists between trajectory solutions 
[of \eq{modnewt1}] and (scattering) TIQM states.\cite{poirier10nowave}

Any fourth-order ODE can be rewritten as a set of four coupled, 
first-order ODE's.  Remarkably, Eq.~(\ref{modnewt1}) can be 
rewritten:
\begin{equation}
\dot{x} = \frac{s}{m}\,\,;\,\,
\dot{p} = - {\partial V(x) \over \partial x}\,\,;\,\,
\dot{s} = \frac{4rs^4}{m\hbar^2}\,\,;\,\,
\dot{r} = \frac{p-s}{m}-\frac{8r^2s^3}{m\hbar^2}\ ,
\end{equation}
which are {\em Hamilton's equations for a 2D system}, 
\begin{equation}
\dot{x}=\frac{\partial H}{\partial p}\,\,;\,\,
\dot{p}=-\frac{\partial H}{\partial x}\,\,;\,\,
\dot{r}=\frac{\partial H}{\partial s}\,\,;\,\,
\dot{s}=-\frac{\partial H}{\partial r}, \label{Hameq}
\end{equation}
for the Hamiltonian
\begin{equation}
 H(x,p,r,s) = \frac{s(2p-s)}{2m} + V(x) - \frac{2r^2s^4}{m\hbar^2}\ .
\label{Ham}
\end{equation}
In the above equations, $(x,p)$ are the ``classical'' dimension
phase space variables, and $(r,s)$ correspond to an additional, 
``quantum'' dimension, essentially describing quantum interference. 
Note that Eq.~(\ref{Ham}) reduces to the classical Hamiltonian 
when $r=0$ and $s=p$. 

In terms of the time derivatives of $x(t)$, 
\begin{equation} 
   s=m \dot{x} \,\,;\,\,
   r=\frac{\hbar^2\ddot{x}}{4m^2\dot{x}^4} \,\,;\,\,
   p = m\dot{x} + \frac{\hbar^2}{4m}\left(\frac{\stackrel{...}{x}}{\dot{x}^4}
        -\frac{2\ddot{x}^2}{\dot{x}^5} \right) \label{PSvar}.
\end{equation}
Substitution of Eq.~(\ref{PSvar}) into Eq.~(\ref{Ham}) then reveals
$H$ to be the conserved Noether energy of Eq.~(\ref{1DE}).
For the free particle case [$\partial V / \partial x=0$], $p$ is a second 
conserved quantity, in involution with $H$
(and also derivable from Noether's theorem). The importance of $p$ is difficult 
to overstate; it represents the ``particle momentum,'' analogous to 
the well-known ``particle energy,'' $E=H$.\cite{holland,wyatt} Yet 
remarkably, $p$ has barely been considered\cite{bouda03} in the previous
literature, which generally regards $s=m \dot x$ as particle 
momentum. Note that $p$ is conserved for {\em all} free particle  
TIQM states, including those exhibiting  interference, whereas 
$s$ is conserved {\em only} for plane wave states---i.e., the classical 
special case for which $p=s$. 

The above Hamiltonian approach is proving extremely useful in 
numerical calculations of quantum reactive scattering 
phenomena.\cite{poirier12ODE} For any 1D TIQM application, {\em exact quantum} 
results are obtained simply by propagating a single $x(t)$ trajectory 
using the  {\em classical} Hamilton's  equations  of \eq{Hameq}, until $p(t)$ flattens
asymptotically to the desired level of numerical accuracy. The final $p$ value then provides a direct 
measure of the quantum reaction probability. Note that Eq.~(\ref{Hameq}) is 
amenable to efficient, symplectic numerical integrators as are used in classical 
simulations, and the conservation of $E$ can  be monitored as 
an on-the-fly measure of computed accuracy. This approach
is extremely robust, accurate, and efficient, leading to 15 digits of accuracy,
even in the extremely deep tunneling regime (where absorbing potentials 
render conventional exact quantum  scattering methods intractible).  
For many-D TIQM applications, classical-like sampling over
quantum trajectory initial conditions  leads to an approximate simulation scheme
that has also proven to be remarkably accurate (i.e., to two or three 
digits).\cite{poirier12ODE}


\section{The 1D Time-dependent Case}
\label{1Dtdse}

For the case of 1D time-dependent quantum mechanics (TDQM), any 
self-contained formulation must involve a PDE, rather than an ODE.  
It is no longer possible to exactly represent a quantum state  as a single
trajectory, $x(t)$, but rather as a one-parameter {\em ensemble} of trajectories, $x(C,t)$, 
where the real-valued, space-like coordinate $C$ labels individual trajectories. 
The equation of motion should be a PDE involving $C$ and $t$ derivatives, 
preferably derived from a field-theoretic action principle.

In Refs.~\onlinecite{poirier10nowave}, \onlinecite{holland05}, and \onlinecite{holland05b}, 
$C$ was chosen as the initial trajectory value [$x(C,0)\! \!= \!\!x_0\!\! =\!\! C$]. 
The resultant PDE is complicated, exhibits explicit $x_0$ 
dependence through the initial probability density, $\rho_0(x_0)\!\! = \!\!\rho(x_0,0)$, and 
bears little resemblance to \eq{modnewt1}. In addition, $Q$ is expressed 
in terms of $C$ rather than $t$ derivatives of $x$.  The PDE can be simplified by a 
better choice of the trajectory parameter, $C$, which  in 
general can be taken to be any monotonic function of $x_0$ (regardless of
the initial wavefunction).  A crucial 
idea of the current paper is that {\em $C$ should be chosen so as to 
uniformize the probability density}. In particular, since 
$\rho_C(C)\, dC = \rho(x,t)\,dx$, if we choose 
\begin{equation}
 C = \int_{-\infty}^{x_0} \rho_0(x_0')\, dx_0'\ ,
\label{Ceq}
\end{equation}
then $C$ takes values from $0$ to $1$ (for normalized wavepackets),
and $\rho_C(C)=1$.
 
Working with \eq{Ceq} (or any uniformizing choice of $C$), and writing $x'=
\partial x /\partial C$,  $\xdot=\partial x / \partial t$ etc., 
the PDE of \Ref{poirier10nowave} simplifies very substantially to 
\begin{equation}
m \xddot  + {\partial V(x) \over \partial x} 
 + \frac{\hbar^2}{4m}\left( 
\frac{x''''}{{x'}^4} - 8 \frac{x'''x''}{{x'}^5} + 10 \frac{{x''}^3}{{x'}^6}
    \right)
=0 \ . 
\label{modnewt2a}
\end{equation} 
Equation~(\ref{modnewt2a}) is the perturbed Newton equation for the 1D 
TDQM case; it has no explicit coordinate dependences. It also bears
an extremely close resemblance to Eq.~(\ref{modnewt1}), obtained by 
replacing $C$ derivatives with $t$ derivatives in the last term on the 
left hand side (representing the quantum force). 
More formally, Eq.~(\ref{modnewt1}) is obtained on looking for travelling wave 
solutions of Eq.~(\ref{modnewt2a}), i.e. solutions of the form 
$x(C,t) = x(t-\lambda C)$ where $\lambda$ is a constant. In the 1D TIQM context,
$t$ thus serves as an effective uniformizing coordinate. 
Equation~(\ref{modnewt2a}) also admits $t$-independent solutions,
which correspond to the {\em bound} (fluxless) 1D TIQM quantum states 
(the previous section concerns only the scattering states). 

As in the 1D TIQM case, Eq.~(\ref{modnewt2a})
is a variational equation, obtained by extremizing the action
\eb
\int \! \! \! \int dC \,dt\, \left [
\frac12 m {\xdot}^2 - V(x) - \frac{\hbar^2}{4m} 
 \left( \frac{x'''}{{x'}^3}  -\frac52 \frac{{x''}^2}{{x'}^4}\right)  \right] , 
\label{action}
\ee
c.f. \eqs{1DL}{Q1D}.
This action is invariant under translations of  both coordinates
$t$ and $C$. By Noether's theorem this gives rise to two conservation laws, 
which are easily found to be, respectively, 
\begin{eqnarray}
{\partial \over \partial t} 
\left[  
 \frac12 m \dot{x}^2  + V(x)  + \frac{\hbar^2}{4m} \left( \frac{x'''}{{x'}^3} - \frac52 \frac{{x''}^2}{{x'}^4} \right)
\right] && \label{Econs}  \\ 
+ \frac{\hbar^2}{4m} {\partial \over \partial C}
\left[
 \left( \frac{x'''}{{x'}^4}-\frac{2{x''}^2}{{x'}^5} \right) \dot{x}  +2\frac{x''\dot{x}'}{{x'}^4} - \frac{\dot{x}''}{{x'}^3}
\right] 
&=&  0 \ , \nonumber \\
{\partial \over \partial t} \left[
m\dot{x}x'  
\right] + 
{\partial \over \partial C} \left[-\frac12m\dot{x}^2 + V(x) +\frac{\hbar^2}{4m}\left(\frac{x'''}{{x'}^3} - \frac52\frac{{x''}^2}{{x'}^4} \right) 
\right] & = &  0 . \label{Ccons}
\end{eqnarray}
The first corresponds to conservation of energy. 
In the free particle case, there is also a momentum 
conservation law, arising from $x$-translation symmetry: 
\begin{equation}
{\partial \over \partial t} \left[ m\dot{x}  \right] + \frac{\hbar^2}{4m}
{\partial \over \partial C}
 \left[ \frac{x'''}{{x'}^4}-\frac{2{x''}^2}{{x'}^5} \right] =  0 \  
 \label{pcons}
\end{equation}

Interpreted as hydrodynamical balance equations of the general form
$\partial A / \partial t + \partial B / \partial C=0$,
the first square bracket in \eq{Econs} [\eq{pcons}] represents the energy 
[momentum] density, and the second term the corresponding flux. 
This designation is only determined up to addition of a $C$-derivative 
to the density $A$ and subtraction of the corresponding $t$-derivative 
from the flux $B$. Consequently, the energy density need not conform to
the standard TDQM ``field'' form,\cite{holland} but instead may be 
chosen to be the particle energy, $T\!\!+\!V\!\!+\!Q$, as in \eq{Econs}.   
This choice is appropriate for the Lagrangian $T\!\!-\!\!V\!\!-\!Q$
[cf. Eqs.~(\ref{1DL}) and (\ref{1DE})], and has the great advantage of being
conserved along individual trajectories in the TIQM limiting case
(in general, only the total ensemble energy is conserved).
However, in the momentum conservation law \eq{pcons}, it does not seem to 
be possible to use a density that reduces to the particle momentum 
$p$ in \eq{PSvar}.

Also of note is the balance equation, Eq.~(\ref{Ccons}).
For any autonomous Euler-Lagrange PDE, the Lagrangian density $L$ is 
determined only up to the addition of a divergence (i.e., the sum of  
$C$- and $t$-derivatives). We have chosen forms of the action 
[Eq.~(\ref{action})] and the conservation laws such that $L$ appears 
as (minus) the flux in Eq.~(\ref{Ccons}). Via gauge transformations, it is 
possible to eliminate the third-order derivative from the Lagrangian and 
energy densities, but not from the flux of Eq.~(\ref{Ccons}).  The choice of 
$L$ we have made has the advantage that 
the trajectory action, 
$S(C,t)=\int_0^t L(C,t') dt'$,  expressed in units of $\hbar$, can be identified 
with the change in phase of $\Psi$.
 
We now consider Gaussian wavepacket evolution under the free particle 
($V=0$) and harmonic oscillator ($V=\frac12 m\omega^2 x^2$) potentials. 
The respective $x(C,t)$ solutions are
\begin{eqnarray}
& x_0+\frac{p_0(t-t_0)}{m} + 
        a\   {\rm erfinv}(2C-1)\sqrt{1 + \frac{\hbar^2(t-t_0)^2}{m^2a^4}}  \qquad  \text{and}  &  \label{sol1} \\
 &  x_0\cos\omega (t-t_0) + \frac{p_0\sin\omega (t-t_0)}{m\omega} + a \ {\rm erfinv}(2C-1) \times 
   \sqrt{ \cos^2\omega(t-t_0) + \frac{\hbar^2 \sin^2\omega(t-t_0)}{m^2a^4\omega^2} }, &
   \label{sol2}
\end{eqnarray}
where $x_0,p_0,t_0,a$ are real wavefunction parameters, and $0\!\le \!C\! \le \!1$.  
Note these solutions diverge as $C\! \rightarrow\! 0$ or $1$.
In general, $x(C,t)$ must diverge at the $C$ endpoints;
modulo this requirement, any solution of Eq.~(\ref{modnewt2a}) (or its arbitrary-$C$
generalization) can be used to reconstruct a normalized solution $\Psi(x,t)$ of the 1D TDSE. 

We have successfully numerically integrated Eq.~(\ref{modnewt2a}) for an 
Eckart  potential and initial Gaussian wavepacket, using the Stormer-Verlet and other
methods.  Initial results compare favorably with those obtained using standard $\Psi$-based 
methods, but further improvements in efficiency and accuracy are planned. In any case, 
these calculations represent a milestone achievement, as the {\em first successful synthetic 
Bohmian quantum trajectory calculations ever achieved for a system with substantial reflection 
interference}---a much-sought goal eluding chemical dynamics researchers 
for over a decade, and a major hurdle preventing exact 
quantum wavepacket calculations for large molecular systems with few reaction
pathways.\cite{wyatt,babyuk06}  


\section{The many-D Time-dependent Case}
\label{manyDtdse}

The 1D analysis generalizes to many-D. The single variable
$x$ is replaced with the $n$-dimensional configuration
space vector ${\bf x}$, with $C$ likewise replaced with ${\bf C}$, 
so that ${\bf x} ({\bf C},t)$ represents an  
$n$-parameter family of trajectories. One option\cite{holland05} is to take 
${\bf C}={\bf x}_0$, though we wish to consider more general choices for which
${\bf C}$ and ${\bf x}_0$ are related via any invertible coordinate transformation.

In analogy with the 1D case, it can be shown that for an arbitrary choice of the 
parametrization ${\bf C}$, 
\begin{equation}
\rho({\bf x},t) = \frac{\rho_{\bf C}({\bf C})}{\det J},
\label{Rdef} 
\end{equation}
where $J$ is the Jacobi matrix, $J^i_{~j}=\partial x^i /\partial C^j$.
As in the 1D case, we mostly work with a uniformizing
${\bf C}$ for which $\rho_{\bf C}({\bf C})=1$.
The resulting perturbed Newton equation, i.e. the many-D 
generalization of \eq{modnewt2a}, can be written in various 
different forms, the most compact being
\begin{equation}
m\, \ddot {x^i} + {\partial V({\bf x}) \over \partial x^i} - \frac{\hbar^2}{4m}
 \frac{\partial}{\partial C^m} \left( 
K^k_{~i}K^m_{~j} \frac{\partial^2 K^l_{~j}}{\partial C^k \partial C^l}
\right) 
   = 0  \ .
\label{f3} 
\end{equation}
Here $K=J^{-1}$ denotes the inverse Jacobi matrix. The Einstein summation convention 
is used, albeit with some mismatched indices as we are 
currently assuming the Euclidean metric on ${\bf x}$ space.

Equation~(\ref{f3}) is the variational PDE for the action
\begin{eqnarray} 
&&\int \! \! \!\int d^nC \, dt  \left[ 
\frac12 m \, \dot{\bf x} \cdot \dot {\bf x} - V({\bf x}) 
-Q \right]\ , \qquad\text{where} \label{f4} \\  
&& Q = - \frac{\hbar^2}{4m} \left( 
K^k_j \frac{\partial^2 K^l_j}{\partial C^l\partial C^k} 
+ \frac12 \frac{\partial K^l_j}{\partial C^l} \frac{\partial K^k_j}{\partial C^k} 
\right)\ .\label{multiDq} 
\end{eqnarray}
Equation~(\ref{f4}) is the many-D generalization of \eq{action}, which preserves
the $L=T\!-\!V\!-\!Q$ and $\Psi$ phase properties. 
Like \eq{action}, \eq{f4} has associated laws of conservation of energy and momentum.
Invariance under $C$ translation is replaced by invariance under the 
infinite-dimensional group of volume preserving diffeomorphisms. 

For a general (not necessarily uniformizing) 
choice of ${\bf C}$, and allowing a non-Euclidean metric $g_{ij}({\bf x})$ on 
${\bf x}$-space 
[note  in this case a factor 
of $\sqrt{g}$ should be inserted in the denominator of the RHS of Eq.~(\ref{Rdef})],
the action, in a simpler gauge involving only second derivatives, becomes
\begin{eqnarray} 
&  & \int \! \! \! \int \rho_{\bf C}({\bf C}) \,d^nC \,dt\,  \left[ \frac12 m \, g_{ij}({\bf x})\, \dot{x^i}\, \dot{x^j} 
 - V({\bf x}) \right.  \nonumber \\ 
 & &  - \left.\frac{\hbar^2}{8m} g^{jl}({\bf x}) \left(\frac{\det J}{\rho_{\bf C}({\bf C})}\right)^2 
(J^{-1})^i_{~j} 
(J^{-1})^k_{~l} 
\frac{\partial}{\partial C^k}\left(\frac{\rho_{\bf C}({\bf C})}{\det J}\right)  
\frac{\partial}{\partial C^i}\left(\frac{\rho_{\bf C}({\bf C})}{\det J}\right) 
\right],\label{f7} 
\end{eqnarray}
cf. Eq. (3.5) in \Ref{holland05b}. Note that a many-D solution of the TDSE 
gives rise to a solution of \eq{f3}, but the converse holds only
if the initial $\bf \xdot$ field is a gradient; this property is 
then preserved by the evolution \eq{f3}.\cite{holland05b}  

\section{CONCLUDING REMARKS}
\label{conclusion}

We have developed a self-contained, trajectory-based formulation
of spin-free nonrelativistic TDQM, achieving all goals as outlined in the 
Introduction. Further developments are underway. Theoretical progress will 
require a correct treatment of spin, relativity, particle indistinguishability, 
and second quantization---with a promising start having been made by Holland 
and others.\cite{holland,holland05,holland05b}
The invariance of \eq{f7} under volume-preserving 
diffeomorphisms suggests a connection with gravity, though the physical significance 
of ${\bf C}$ is not yet clear. 

Numerically, the prospect of stable, synthetic quantum trajectory calculations 
for many-D molecular applications will be fully explored, as the benefits here
could prove profound.\cite{wyatt,poirier10nowave,poirier12ODE} Our formalism offers 
flexibility for restricting action extremization to trajectory ensembles 
of a desired form (e.g., reduced dimensions), thereby providing useful 
variational approximations. Our exact TDQM equations are PDEs, not single-trajectory 
ODEs---the entire ensemble must be determined at once.  But they provide the great 
advantage of making no reference to any external fields, such as densities or
wavefunctions.  Alternatively, the many-D Hamiltonian ODE approach is 
approximate, but evidently quite accurate.\cite{poirier12ODE}

Regarding interpretation, we draw no definitive conclusions here. However,  it is clearly
of great significance that the form of $Q$ can be expressed in 
terms of $x$ and its $C$  derivatives---implying the key idea that the interaction of nearby 
{\em trajectories}, rather than particles, is the source of all empirically observed quantum 
phenomena (suggesting a kind of ``many worlds'' theory, albeit one very different from
\Ref{everett57}). As such, it is locality in {\em configuration} space, rather than in the usual 
position space {\em per se}, that is relevant. In effect, we have a hidden variable theory
that is local in configuration space, but nonlocal in position space---though the latter
is hardly ``spooky'' in the present nonrelativistic context [even classical theory is
nonlocal in this sense, depending on $V(\bf x)$].  Many ramifications are anticipated
for a wavefunction-free interpretation of measurement, entanglement, etc. 
One wonders whether Bohm would have abandoned pilot waves, had he known
such a formulation was possible---or, for that matter, whether the 
notion of quantum trajectories might  have actually appealed to Einstein.

\section{ACKNOWLEDGEMENTS}
This work was supported by grants from the Robert A. Welch Foundation 
(D-1523) and the National Science Foundation (CHE-0741321).  Travel 
support from the US--Israel Binational Science Foundation (BSF-2008023) is also acknowledged.

%
%

\end{document}